\documentclass[preprint,aps,nofootinbib]{revtex4}

\newcommand {\bc}{\begin{center}}
\newcommand {\ec}{\end{center}}
\newcommand {\bea}{\begin{eqnarray}}
\newcommand {\eea}{\end{eqnarray}}
\newcommand {\be}{\begin{equation}}
\newcommand {\ee}{\end{equation}}

\def\lsim{\mathrel{\rlap{\lower4pt\hbox{$\sim$}}
    \raise1pt\hbox{$<$}}}               
\def\gsim{\mathrel{\rlap{\lower4pt\hbox{$\sim$}}
    \raise1pt\hbox{$>$}}}  
\usepackage{graphicx}

\begin{document}


\title{Dissipative superfluid hydrodynamics for the unitary Fermi gas}

\author{Jiaxun Hou and Thomas Sch{\"a}fer}

\affiliation{Department of Physics, North Carolina State University,
Raleigh, NC 27695}

\begin{abstract}
In this work we establish constraints on the temperature dependence 
of the shear viscosity $\eta$ in the superfluid phase of a dilute Fermi 
gas in the unitary limit. Our results are based on analyzing experiments 
that measure the aspect ratio of a deformed cloud after release from 
an optical trap. We discuss how to apply the two-fluid formalism to 
the unitary gas, and provide a suitable parametrization of the equation
of state. We show that in expansion experiments the difference between
the normal and superfluid velocities remains small, and can be treated
as a perturbation. We find that expansion experiments favor a shear
viscosity that decreases significantly in the superfluid regime. Using 
an exponential parametrization we find $\eta(T_c/(2T_F))\lsim 0.37\,
\eta(T_c/T_F)$, where $T_c$ is the critical temperature, and $T_F$ 
is the local Fermi temperature of the gas.
\end{abstract}

\maketitle
\section{Introduction}
\label{sec_intro}

 The dilute Fermi gas at unitarity has emerged as an important example 
of a strongly correlated quantum fluid \cite{Bloch:2007,Giorgini:2008}. 
Measurements of equilibrium and non-equilibrium properties provide 
important benchmarks for a variety of physical systems, ranging from 
dilute neutron matter in neutron stars to the quark gluon plasma probed 
in relativistic heavy ion collisions. The unitary Fermi gas is a
system of non-relativistic spin-1/2 particles interacting via an
interaction of zero range tuned to infinite scattering length. This 
means that the system is strongly interacting, but the only scales 
in the problem are those that can be defined in the non-interacting 
gas, for example the Fermi momentum $k_F=(3\pi^2n)^{1/3}$, where $n$
is the density of the gas. From the Fermi momentum we can construct
the Fermi energy $E_F=k_F^2/(2m)$, and the Fermi temperature $k_BT_F=
E_F$, where $k_B$ is the Boltzmann constant. As an example, consider
the superfluid transition in a unitary Fermi gas. Based on dimensional
analysis, the critical temperature must be proportional to the local 
Fermi temperature, $T_c\sim T_F$. Indeed, experiments find $T_c=0.167(3)
T_F$ \cite{Ku:2011}.

  A remarkable property of the unitary Fermi gas is nearly perfect
hydrodynamic flow \cite{Schafer:2009dj,Adams:2012th,Schaefer:2014awa,
Schafer:2007pr,Turlapov:2007}, originally discovered in \cite{OHara:2002}.
In this experiment, the authors observed nearly ideal flow in a dilute
unitary Fermi gas after release from a deformed trap. The deformation
of the trap implies that pressure gradients in the short (transverse) 
direction of the trap lead to preferential acceleration in this direction. 
As a consequence, the aspect ratio of the cloud changes from being
elongated in the longitudinal direction to being elongated in the transverse
direction. Viscosity counteracts this behavior, and detailed studies of
the time evolution can be used to extract the shear viscosity as a 
function of $T/T_F$. 

 Shear viscosity is a dimensionful quantity, and it is natural to 
consider the dimensionless ratios $\eta/n$ or $\eta/s$, where $s$
is the entropy density. We have also set $\hbar=k_B=1$, where $\hbar$
is Planck's constant, and $k_B$ is Boltzmann's constant. Previous 
analysis finds that for $T>T_F$ the shear viscosity of the gas can be 
understood in terms of kinetic theory, and that in this regime $\eta/n
>1$ \cite{Bluhm:2015bzi}. Near the critical temperature the shear 
viscosity enters the quantum regime, $\eta/n<1$, and the viscosity
is only weakly temperature dependent. At $T_c$, we find $\eta/n\simeq
0.4$ \cite{Bluhm:2017rnf}, in agreement with calculations based on
the Kubo relation and resummed perturbation theory \cite{Enss:2010qh},
see also \cite{Hofmann:2011qs,Hofmann:2019jcj,Frank:2020oef}.

Our main goal in the present work is to constrain the behavior of $\eta/n$ 
below the superfluid phase transition using existing expansion experiments,
in particular the results of Joseph et al.~\cite{Joseph:2014}. Several, 
qualitatively different, predictions and results regarding the behavior of 
$\eta/n$ below $T_c$ can be found in the literature. In Ref.~\cite{Rupak:2007vp} 
we argued that for $T\ll T_c$ kinetic theory in terms of phonon quasi-particles 
is reliable, and that it predicts that $\eta/n$ grows rapidly as $T/T_F\to 0$. 
On the other hand, Ref.~\cite{Guo:2010} proposed a different quasi-particle 
model, which predicts $\eta/n\to 0$ as $T\to 0$. This behavior is seen in the 
Quantum Monte Carlo calculation of \cite{Wlazlowski:2013owa}, and in the 
simplified experimental analysis in \cite{Joseph:2014}. A recent experimental 
study of sound attenuation in a unitary Fermi gas finds that sound diffusivity 
is approximately constant below $T_c$ \cite{Patel:2019}. Different behaviors
are also seen in the two isotopes of helium. In $^4$He the viscosity is 
dominated by rotons and phonons. It is approximately constant near $T_c$,
and grows very steeply as $T\to 0$ \cite{Schafer:2009dj}. In $^3$He, on the 
other hand, the viscosity shows a steep drop below $T_c$ \cite{Guo:2010,
Vollhardt:1990}.

 The basic tool for analyzing the expansion experiment below $T_c$ is 
superfluid (two-fluid) hydrodynamics. When solving the two-fluid equations 
in an expanding system, careful attention has to be paid to the frame 
dependence of the equations of motion. We review this issue in Section
\ref{sec_sfl_Landau} and \ref{sec_con}. Section ~\ref{sec_ufg} discusses
simplifications that appear if the fluid is scale invariant, as is the case 
for the unitary Fermi gas. Sections \ref{sec_sfl_Landau}-\ref{sec_ufg} can 
be skipped if the reader is primarily interested in the analysis of the 
experiments of Joseph et al.~\cite{Joseph:2014}. Simple solution of the
two-fluid equations are discussed in Sect.~\ref{sec_sol}. We discuss an
approach based on treating the difference between the superfluid and normal
velocities as a small parameter, and present an analysis of the experimental
data using this method in Sect.~\ref{sec_analysis}. We conclude in
Sect.~\ref{sec_outlook}, and details regarding the equation of state and 
the initial conditions for two-fluid hydrodynamics are discussed in 
several appendices. 

\section{Superfluid hydrodynamics}
\label{sec_sfl_Landau}

 Fluid dynamics is based on conservation laws, combined with approximate
local thermodynamic equilibrium. Thermodynamic relations are encoded in 
an equation of state, which can be measured in a fluid at rest in the 
laboratory frame. In this section we review how this information is used 
in the fluid dynamic description of an expanding fluid. For simplicity, 
consider first a normal fluid, for example the unitary Fermi gas above 
$T_c$. There are five hydrodynamic variables, the mass density $\rho$, the 
energy density ${\cal E}$ (or, alternatively, the entropy density $s$), 
and the mass current $\vec\jmath$. Note that the mass current is equal 
to the density of momentum, so that the total momentum of the fluid is 
the integral of $\vec{\jmath}$ over the volume occupied by the fluid. 

 For a fluid element centered at position $\vec{x}$ with mass current
$\vec\jmath\,(\vec{x})$ there is a Galilean boost with boost velocity
$\vec{v}=\vec\jmath/\rho$ that transforms the conserved charges into 
the local rest frame of the fluid, defined by $\vec\jmath\,(\vec{x})=0$. 
In this frame the energy density of the fluid is ${\cal E}_0= {\cal E}
-\vec{\jmath}^{\;2}/(2\rho)$, see App.~\ref{sec_gal}. The energy density 
in the rest frame satisfies the standard thermodynamic identity
\be 
d{\cal E}_0=\mu_0dn + Tds, 
\ee 
where $\mu_0$ is the  chemical potential in the fluid rest frame and 
$n=\rho/m$ is the particle number density. The energy in the laboratory 
frame satisfies
\be
\label{dE_nfl}
d{\cal E} = \left(\mu_0-\frac{1}{2}m\vec{v}^{\,2}\right)
   dn + Tds + \vec{v}\cdot d\vec\jmath \, .
\ee
The pressure is given 
by the Legendre transform of equ.~(\ref{dE_nfl}) with respect to 
$n,s$ and $\vec{\jmath}$. We obtain $P=\mu_0 n+sT-{\cal E}_0$, and 
the pressure satisfies the Gibbs-Duhem relation $dP=nd\mu_0 + sdT$. 

  In a superfluid there are three additional hydrodynamic variables, 
the components of the  superfluid velocity $\vec{v}_s$. As a result, there 
no longer is a unique local frame in which the fluid is at rest. In this section
we will follow Landau \cite{Landau-Hydro} and consider thermodynamics in
the restframe of the superfluid. The final result is of course independent
of the choice of frame, and we describe thermodynamic relations in the
rest frame of the normal fluid in App.~\ref{sec_sfl_nfl}. The mass current
in the frame of the superfluid is $\vec{\jmath}_0= \vec\jmath-\rho \vec{v}_s$.
Using Galilean invariance we can determine the energy density in the
superfluid rest frame
\be 
\label{E_sfl}
{\cal E}_s = {\cal E} -\vec\jmath_0\cdot\vec{v}_s -\frac{1}{2}\rho v_s^2 \, , 
\ee
We now view the energy density in the superfluid frame as a function of
$\vec{\jmath}_0$, 
\be 
\label{dEs}
d{\cal E}_s = \mu_s dn + Tds + \vec{w}\cdot d\vec{\jmath}_0\, , 
\ee
where we have defined $\mu_s$, the chemical potential in the superfluid frame, 
and the velocity $\vec{w}$. For the energy in the lab frame 
equ.~(\ref{dEs}) implies
\be 
\label{dE_sfl}
 d{\cal E} = \mu_{\jmath_0} dn + Tds 
   + \vec{v}_n\cdot d\vec{\jmath}_0\, 
   + \vec\jmath \cdot d\vec{v}_s\, , 
\ee
where
\be
\mu_{\jmath_0}=\mu_s+\frac{1}{2}m v_s^2\, , 
\ee
and we have defined $\vec{v}_n=\vec{w}+\vec{v}_s$, the velocity of the normal 
fluid. We note that $\vec{w}=\vec{v}_n-\vec{v}_s$ is Galilei invariant.
We can perform a Legendre transformation with respect to $n,s$ and
$\jmath_0$. We obtain the pressure 
\be
\label{P_sfl}
 P = -{\cal E}_s + \mu_sn + Ts + \vec\jmath_0\cdot\vec{w}\, ,
\ee
and the Gibbs-Duhem relation
\be
\label{P_GD_sfl}
dP = n d\mu_s + sdT + \vec\jmath_0\cdot d\vec{w}\, .  
\ee
Based on Galilean invariance we can write $\vec\jmath_0=\rho_n \vec{w}$,
which defines the normal fluid density $\rho_n$, as well as the superfluid
density $\rho_s=\rho-\rho_n$. This definition leads to the standard two-fluid
relation $\vec{\jmath} =\rho_n\vec{v}_n + \rho_s\vec{v}_s$. We note that
\be
\label{rho_n_def}
 \rho_n = 2\left. \frac{\partial P}{\partial w^2}\right|_{\mu_s,T}, 
\ee
and $\rho_n\geq0$ implies that at fixed $\mu_s$ the pressure increases 
with $|\vec{w}|$. Finally, we note that 
\be 
\label{P_sfl_2}
 P + {\cal E} =  \mu_s n + Ts + \rho_n\vec{w}\cdot\vec{v}_n
   + \frac{1}{2} \rho v_s^2\, ,
\ee
which will be useful in the following section. 

\section{Conservation Laws}
\label{sec_con}

As in ordinary fluid dynamics there are conservation laws
for the mass density, the momentum density, and the energy
density of the fluid. Mass conservation is
\be
\label{m_cons}
\partial_t \rho + \vec\nabla\cdot\vec\jmath = 0\, , 
\ee
where $\rho=\rho_n+\rho_s$. The particle density is decomposed 
analogously, $n=n_n+n_s$. The mass equation does not receive any 
dissipative corrections. Momentum conservation is
\be
\label{j_cons}
\partial_t \jmath_i + \nabla_j \Pi_{ij} = 0\,
\ee
where the stress tensor $\Pi_{ij}$ can be split into an ideal
and a dissipative part, $\Pi_{ij}=\Pi^{(0)}_{ij}+\delta\Pi_{ij}$.
The ideal part is 
\be 
\label{Pi_ideal}
 \Pi_{ij}^{(0)}= P\delta_{ij} + \rho_n(v_n)_i(v_n)_j
     + \rho_s (v_s)_i(v_s)_j\, . 
\ee
Note that the stress tensor in the superfluid rest frame is 
$\Pi_{ij}^{(0,s)}= P\delta_{ij} + \rho_n w_i w_j$. This expression
follows from Galilean invariance and the second law of thermodynamics. 
The transformation of $\Pi_{ij}$ under Galilean boosts is given
in equ.~({\ref{Pi_boost}). The dissipative terms are given by 
\be 
\label{Pi_diss}
 \delta \Pi_{ij} = -\eta \sigma_{ij}
 -\zeta_2\delta_{ij} \left(\vec\nabla\cdot \vec{v}_n\right) 
 -\zeta_1 \vec\nabla\cdot\left(\rho_s\vec{w}\right) \, , 
\ee
where $\eta$ is the shear viscosity, the shear tensor is defined by
\be 
\label{strain}
\sigma_{ij} = \nabla_i(v_n)_j + \nabla_j(v_n)_i 
  -\frac{2}{3}\delta_{ij} (\vec\nabla\cdot\vec{v}_n) \, , 
\ee
and $\zeta_{1,2}$ are bulk viscosities. In a scale invariant fluid both $\zeta_1$ 
and $\zeta_2$ vanish. Energy conservation is 
\be 
\label{e_cons}
\partial_t {\cal E} + \vec\nabla\cdot\vec{Q} = 0
\ee
with $\vec{Q}=\vec{Q}^{(0)}+\delta\vec{Q}$. The ideal energy current 
in the superfluid rest frame is proportional to $\vec{w}$. Using the 
second law of thermodynamics one can show that $\vec{Q}^{(0,s)}=
({\cal E}_s+P-n_s\mu_s)\vec{w}$, and the energy current in the 
lab frame is 
\be 
\label{q_ideal}
 \vec{Q}^{(0)} = \left( \frac{\mu_s}{m} + \frac{v_s^2}{2} \right)\vec\jmath
 + sT \vec{v}_n
 + \rho_n \left( v_n^2 - \vec{v}_n\cdot\vec{v}_s\right)
  \vec{v}_n \, .
\ee
We observe that with the help of equ.~(\ref{P_sfl_2}) the energy current
$\vec{Q}^{(0)}$ can be expressed as 
\be 
\label{q_ideal_2}
 \vec{Q}^{(0)} = \vec{v}_n \left( {\cal E}+P \right) 
   - \rho_s \vec{w} \left( \frac{\mu_s}{m} + \frac{v_s^2}{2} \right) \, . 
\ee
The dissipative correction is 
\be
\label{Q_diss}
\delta Q_i=-\kappa\nabla_i T +\delta \Pi_{ij}(v_n)_j 
   - \delta\chi \rho_s w_i \, , 
\ee
where $\kappa$ is the thermal conductivity, and $\delta\chi$ is 
given in equ.~(\ref{chi_diss}). Finally, superfluid hydrodynamics 
requires an equation of motion for the superfluid velocity. We have
\be 
\label{vs_eom}
 \partial_t \vec{v}_s + \vec\nabla \chi = 0 
\ee
with $\chi=\chi^{(0)}+\delta\chi$ and 
\be 
\label{chi_0}
 \chi^{(0)} = \frac{\mu_s}{m} + \frac{v_s^2}{2}\, .
\ee
The dissipative term is 
\be 
\label{chi_diss}
 \delta \chi = \zeta_4 \vec\nabla\cdot\vec{v}_n
               +\zeta_3 \vec\nabla \left( \rho_s\vec{w}\right) \, . 
\ee
In a scale invariant gas $\zeta_4$ vanishes \cite{Son:2005tj}, but $\zeta_3$ 
is expected to be non-zero.

\section{Unitary Fermi Gas}
\label{sec_ufg}

In a normal fluid the pressure is a function of two variables, 
$P=P(\mu_0,T)$. In a scale invariant fluid we can write
\be 
\label{P-uni-0}
P(\mu_0,T)= m^{3/2}\mu_0^{5/2} p(T/\mu_0).
\ee
In general, the function $p(T/\mu)$ has to be determined from
experiment. A parametrization of $p(T/\mu)$ for the dilute Fermi 
gas at unitarity can be found in \cite{Bluhm:2017rnf}, and in 
App.~\ref{sec_eos}. In a superfluid the pressure is a function 
of three variables, $P(\mu_s,T,\vec{w})$. Using scale invariance 
we can define a function $p_s$ of two variables, 
\be
\label{P-uni-3}
P(\mu_s,T,\vec{w}) = m^{3/2}\mu_s^{5/2}
p_s\left( \frac{T}{\mu_s},\frac{mw^2}{\mu_s}\right)\, , 
\ee
and for small $w$ we can expand
\be
\label{P-uni-4}
P(\mu_s,T,\vec{w}) = P_0(\mu_s,T)+P_1(\mu_s,T)w^2 = 
      m^{3/2}\mu_s^{5/2}p_{s0}\left( \frac{T}{\mu_s}\right)
    + m^{5/2}\mu_s^{3/2} w^2 p_{s1}\left(\frac{T}{\mu_s}\right)\, . 
\ee
In a similar fashion, we can expand the density, $n=n_0+n_1 w^2$,
and the entropy density, $s=s_0+s_1 w^2$. Note that for $\vec{w}=0$ 
we have $\mu_s=\mu_0$ so that the function $p_{s0}$ is determined
by the pressure of a fluid at rest, $p_{s0}(T/\mu)=p(T/\mu)$. The
function $p_{s1}$ is related to the normal fluid density
\be 
\rho_n = 2m^{5/2}\mu_s^{3/2} p_{s1}\left(\frac{T}{\mu_s}\right)\, . 
\ee
In fluid dynamics we have to determine the pressure using the 
values of the conserved charges. Consider first a normal
scale invariant fluid. Given the energy density we can determine
the energy density in the rest frame, ${\cal E}_0={\cal E}-
\jmath^2/(2\rho)$. Using the Gibbs-Duhem relation we get
\be 
{\cal E}_0= \left\{ \mu_0\frac{\partial}{\partial \mu_0}
 + T\frac{\partial}{\partial T} - 1 \right\} P(\mu_0,T)\, . 
\ee
Using the universal form of the equation of state in 
equ.~(\ref{P-uni-0}) this implies $P=\frac{2}{3}{\cal E}_0$
and
\be
\label{P_uni_hydro}
P = \frac{2}{3}\left\{ {\cal E}- \frac{\jmath^2}{2\rho}\right\}\,  .
\ee
We observe that the pressure can be determined without using the 
explicit form of the function $p(T/\mu)$. Note that if the fluid
is not scale invariant, then we have to tabulate the equation
of state in the form $P=P({\cal E}_0,\rho)$ in order to 
determine the pressure. We also note that transport coefficients
are functions of $T/\mu_0$. The determination of $T/\mu_0$ 
requires explicit knowledge of the function $p(T/\mu)$. A procedure 
for extracting $T/\mu_0$ was proposed in \cite{Bluhm:2017rnf}. 
Consider the dimensionless ratio
\be
x= \frac{2}{(2\pi)^{3/2}}\frac{(mP)^{3/2}}{n^{5/2}}\, .
\ee
The quantity $x$ is Galilean invariant, and purely a function of 
the inverse fugacity $\zeta = \exp(-\mu_0/T)$. The function $\zeta(x)$ 
can be determined from the function $p(T/\mu)$ defined above. Once 
$\zeta$ is determined we can compute the temperature from
\be
T = G(x) \frac{P}{n}
\ee
where the dimensionless function $G(x)$ is also determined by $p(T/\mu)$,
see App.~\ref{sec_eos}. Once $\zeta$ and $T$ are given, then the chemical 
potential is determined by $\mu = -T\log(\zeta)$. 

  We can now study the analogous problem in the superfluid phase.
We first note that given the local energy density ${\cal E}$ we 
can compute ${\cal E}_s$ using equ.~(\ref{E_sfl}). This calculation
only requires the hydrodynamic variables $\rho,\vec{\jmath}$ and 
$\vec{v}_s$. Furthermore, equ.~(\ref{P_sfl}) and (\ref{P_GD_sfl}) 
imply that 
\be 
{\cal E}_s= \left\{ \mu_s\frac{\partial}{\partial \mu_s}
 + T\frac{\partial}{\partial T} 
 +\vec{w}\frac{\partial}{\partial\vec{w}}
 - 1 \right\} P(\mu_s,T,w) \, . 
\ee
Using the equation of state of the unitary gas at $O(w^2)$ we 
get
\be 
P = \frac{2}{3} \left\{ {\cal  E}_s   
 -\frac{1}{2}\vec\jmath_0 \cdot\vec{w}  \right\}\, . 
\ee
This result is more difficult to use than equ.~(\ref{P_uni_hydro}) 
because, whereas ${\cal E}$, ${\cal E}_s$, and $\jmath_0$ are (primary)
hydrodynamic variables, $\vec{w}$ is determined by the equation 
of state, $\vec{w}=\vec\jmath_0/\rho_n$ with $\rho_n=\rho_n
(\rho,{\cal E}_s,\jmath_0)$. One possible approach is to tabulate
the function $\rho_n(\rho,{\cal E}_s,\jmath_0)$ for the equation
of state given in equ.~(\ref{P-uni-3}). Another option is to 
solve for $\rho_n$ and $P$ perturbatively in $w^2/\mu_s$. In the
perturbative approach we set\footnote{
Note that $P(\{w^0\})$ is equal to the exact pressure up to 
errors of order $w^2$, but is is different from the pressure 
in the limit $w\to 0$, which we denoted by $P_0$ in 
equ.~(\ref{P-uni-4}).} 
\be
P(\{w^0\})= \frac{2}{3}{\cal E}_s
\ee
and define
\be
x(\{ w^0 \}) = \frac{2}{(2\pi)^{3/2}}
  \frac{(mP(\{ w^0\})^{3/2}}{n^{5/2}}\, .
\ee
Here, $P(\{w^0\})$ denotes the pressure at order $w^0$, that is 
the exact pressure up to corrections of order $O(w^2)$. We can  
compute the inverse fugacity at this order in the expansion, 
$\zeta\simeq \zeta (x(\{ w^0 \}))$, and the result determines 
the normal fluid fraction at $O(w^0)$
\be 
\left(\frac{\rho_n}{\rho} \right)_{\{w^0\}} 
  = \left. \frac{2p_{s1}(x)}{\frac{5}{2}p_{s0}(x)-x p_{s0}'(x)} 
     \right|_{x=x(\{w^0\})} \, . 
\ee
This result can now be used to compute the pressure at 
$O(w^2)$,
\be
P(\{w^2\})= \frac{2}{3}\left\{ {\cal E}_s
 -    \left(\frac{\jmath^2_0}{2\rho}\right) 
 \left(\frac{\rho}{\rho_n} \right)_{\{w^0\}} 
 \right\} \, . 
\ee
If needed, these results can be used to compute $\mu_s$
and $T$ at order $w^2$.

\section{Simple solutions of the two-fluid equations}
\label{sec_sol}

 It is interesting to note that there are some simple
solutions to the equations of superfluid hydrodynamics that
are relevant to trapped atomic gases. We first observe that
the solution of the hydrostatic equation carries over directly
from the normal fluid case. Consider a fluid confined by an 
external potential $V_{\it ext}(x)$. A static solution of 
the fluid dynamic equations is given by 
\be 
\label{n_LDA}
 n(\vec{x}) = n(\mu_s(\vec{x}),T)\, \hspace{1cm}
 \mu_s(\vec{x})=\mu_c - V_{\it ext}(\vec{x})
\ee
with $\vec{v}_n=\vec{v}_s=0$. This follows directly from the 
Gibbs-Duhem relation $\vec{\nabla} P = n\vec{\nabla}\mu_s$
for $T={\it const}$ and $\vec{w}=0$. We are mostly interested 
in approximately harmonic potentials of the form $V_{\it ext}
(\vec{x}) = m \omega_i^2x_i^2/2$. A number of authors have considered 
small oscillations around the hydrostatic case, see, for example
\cite{Zaremba:1999,Stringari:2013}. The solutions are analogous 
to first and second sound modes in an infinite system.

 In the normal fluid case there is a simple exact scaling solution
to the Euler equation that describes the expansion after a confining
harmonic potential is turned off. For this solution the density expands 
by a scale transformation, $n(x_i,t)=n(x_i/b_i(t),t\!=\!0)$, where 
$n(x_i,0)$ is a solution of the hydrostatic equation. The velocity 
field is a Hubble flow $v_i(\vec{x},t)= \alpha_i(t)x_i$ (no sum over 
$i$), where $\alpha_i(t)=\dot{b}_i/b_i$. The temperature $T$ is only 
a function of time, but not of position. The equation of motion for 
$b_i(t)$ and $T(t)$ is reviewed in \cite{Schaefer:2009px,Schafer:2010dv}.

 We can ask whether this solution generalizes to a solution of 
superfluid hydrodynamics in which the normal and superfluid 
components move together, $\vec{v}_n=\vec{v}_s=\vec{v}$, where 
$\vec{v}$ is the velocity of a normal fluid satisfying the Euler
equation with the equation of state $P(\rho,{\cal E})=P(\rho,{\cal E}_s,
\vec{w}\!=\! 0)$. This is indeed the case. 

 First we note that the momentum density is $\vec\jmath= \rho_n
\vec{v}_n+\rho_s\vec{v}_s=(\rho_n+\rho_s)\vec{v}=\rho\vec{v}$.
This implies that if $\vec\jmath = \rho\vec{v}$ satisfies the 
continuity equations, so does mass current in superfluid hydrodynamics.
The same argument applies to the stress tensor.  For $\vec{v}_n=
\vec{v}_s$ the stress tensor given in equ.~(\ref{Pi_ideal}) assumes the 
normal fluid form $\Pi_{ij}^{(0)} = P\delta_{ij}+\rho v_iv_j$. As a 
consequence, momentum conservation, equ.~(\ref{j_cons}), is satisfied.
Finally, we can study the energy current. If $\vec{w}=0$ then 
equ.~(\ref{q_ideal_2}) reduces to $\vec{Q}^{(0)}=\vec{v}({\cal E}
+P)$, which is the normal fluid form. 

 In superfluid hydrodynamics there is one additional equation, 
which is the equation of motion for the superfluid velocity, 
equ.~({\ref{vs_eom}). The superfluid is accelerated by gradients
of $\mu_s$, not gradients of $P$, and equ.~(\ref{P_GD_sfl}) implies 
that 
\be
\label{P_GD_sfl_2}
  \vec\nabla\mu_s  = \frac{1}{n}
 \left(\vec\nabla P-s\vec\nabla T -
      \rho_n w_i\vec\nabla w_i\right)\, .  
\ee
We conclude that the equation of motion for $\vec{v}_s$ follows
from the Euler equation provided $\vec{w}=0$ and $\vec\nabla T=0$.
As explained above, these conditions are satisfied for the solution
of the Euler equation in an expanding cloud. 

 The correspondence between solution of one and two-fluid hydrodynamics
does not extend to the dissipative case, except in very special 
circumstances. For $\vec{v}_n=\vec{v}_s=\vec{v}$ the dissipative contribution
to the stress tensor, equ.~(\ref{Pi_diss}), is given by 
\be 
\label{Pi_diss_2}
 \delta \Pi_{ij} = -\eta \left( \nabla_i v_j 
   + \nabla_jv_i -\frac{2}{3}\delta_{ij} (\vec\nabla\cdot\vec{v})\right) 
 -\zeta_2\delta_{ij} \left(\vec\nabla\cdot \vec{v}\right)  \, , 
\ee
which is the same as the one-fluid expression (for $\zeta=\zeta_2$). 
However, the viscous correction to the energy current, $\delta Q_i
=\delta\Pi_{ij}v_j$ (see equ.~(\ref{q_ideal_2})), leads to viscous heating 
and a non-zero temperature gradient unless the functional form of the 
viscosity $\eta(n,T)$ is very specially chosen. This implies that we no
longer have $\vec\nabla P = n\vec\nabla\mu_s$, and the solution 
$\vec{v}_n=\vec{v_s}$ is not consistent. 

 However, given that viscous corrections are small we can treat 
$\vec{\nabla}T$ and $\vec{w}$ as perturbations, and solve for 
$\vec{w}$ at leading order. Using equ.~(\ref{m_cons}-\ref{q_ideal})
we find
\be 
\label{w_eom_lin}
 \left(\frac{\partial}{\partial t}+\vec{v}\cdot\vec\nabla\right)
  \vec{w} = -\frac{s}{\rho_n}\vec\nabla T
 + O(w^2)\, . 
\ee
In the absence of a background flow, $\vec{v}=0$, this equation is well 
known from the study of small oscillations in a superfluid, where it 
describes the restoring force in a second sound mode \cite{Landau-Hydro}. 
We observe that the result remains valid in a non-trivial background flow 
$\vec{v}_n\simeq \vec{v}_s \neq 0$, provided the advection of $\vec{w}$ 
in the background flow is taken into account. 

\section{Hydrodynamic analysis in the small $w$ limit}
\label{sec_analysis}

 In this section we discuss an analysis of the data taken in the 
superfluid regime by Joseph et al.~\cite{Joseph:2014}. The same 
data were previously analyzed in the normal fluid regime
in \cite{Bluhm:2017rnf}. In the experiments the gas is released 
from a harmonic trap $V_{\it ext}=\frac{1}{2}m\omega_i^2x_i^2$ 
with trap frequencies $(\omega_x,\omega_y,\omega_z)=(2\pi)(2210,
830,64.3)$ {\it Hz}. After the optical trap is turned off there 
is a residual magnetic bowl characterized by $\omega_{\it mag}=2\pi
\cdot 21.5\,{\it Hz}$. The central temperature of the cloud varies
between $T=(0.05-1.10)T_F$.


  Expansion experiments measure the aspect ratio of the cloud after
the gas is released from the harmonic trap. Hydrodynamic flow develops
because the pressure gradients in the initial configuration accelerate
the gas. If the initial trap is deformed differences in the pressure
gradients in different directions cause the expansion to be fastest
in the short direction of the trap. This phenomenon is known as 
elliptic flow. 

  Joseph et al.~\cite{Joseph:2014} measure the ratio $A_R\equiv
\sigma_x/\sigma_y$, where $\sigma_x$ and $\sigma_y$ are Gaussian fit 
radii in the $x$ and $y$ direction obtained from two-dimensional 
absorption images of the cloud. For the trap configuration studied 
in the experiment this ratio evolves more quickly than $\sigma_x/
\sigma_z$ or $\sigma_y/\sigma_z$. Fig.~\ref{fig:Ar-LowT} shows 
the dependence of $A_R(t^*)$ at a fixed time $t^*=1.2\,{\it msec}$
on the initial temperature of the cloud. Note that $A_R(0)\sim 0.37$,
and the measured values $A_R(t^*)>1$ reflect the elliptic flow
phenomenon discussed above. The main idea of the experiment is 
that shear viscosity counteracts the rise in $A_R$ as a function
of time, and that the dependence of $A_R(t^*)$ on $T/T_F$ constrains
the dependence of shear viscosity on temperature and density.

\begin{figure}[t]
\centering
\includegraphics[scale=0.9]{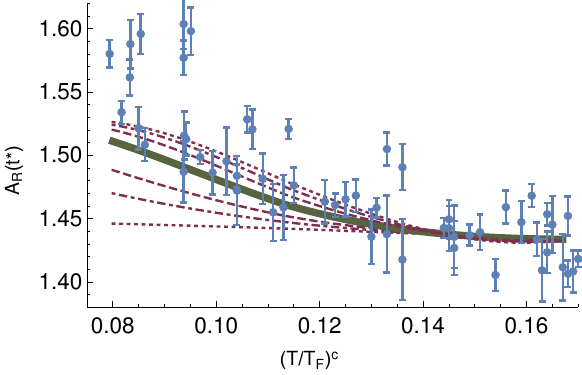}
\caption{\label{fig:Ar-LowT}
Aspect ratio $A_r$ as a function of the temperature $T/T_F$ in units 
of the Fermi temperature at the center of the trap. The data show 
$A_R$ at $t^{*}=1.2 \times 10^{-3}$s after release from the trap
\cite{Joseph:2014}. The lines show the prediction of viscous 
hydrodynamics for different values of the dimensionless parameter 
$\alpha=(0,0.5,1,2,3,4,5)$ (from bottom to top) defined in 
equ.~(\ref{eta_lowT}). The thick green line corresponds to $\alpha=2$. }
\end{figure}

 In our previous work we analyzed the data above $T_c$ assuming 
an expansion of the viscosity in the dimensionless diluteness 
$n\lambda^3$ of the gas, $\eta(n,T)\simeq \eta_{\it vir}(n,T)$ with 
\be
\label{eta_vir}
\eta_{\it vir} (n,T) = \eta_0 (mT)^{3/2} 
  \left\{ 1 + \eta_2 \left(n\lambda^3\right)
            + \eta_3 \left(n\lambda^3\right)^2 + \ldots \right\}.
\ee
Here, $\lambda=[2\pi/(mT)]^{1/2}$ is the thermal de Broglie wavelength.
In \cite{Bluhm:2017rnf} we obtained
\be
\eta_0 =  0.265 \pm 0.02 \,  ,\hspace{1cm}
\eta_2 =  0.060 \pm 0.02  \, . 
\ee
We also found that $\eta_3$ is consistent with zero. Note that $\eta_0$
can be compared to the kinetic theory result $\eta_0=15/(32\sqrt{\pi})
\simeq 0.264$ \cite{Bruun:2005}. Here, we will study whether the data 
constrain the behavior below $T_c$. We consider the parametrization 
\be 
\label{eta_lowT}
 \eta(n,T<T_c(n)) = \eta_{\it vir}(n,T)
   \exp\left( \alpha \left[ 1-\frac{T_c(n)}{T}\right]\right),
\ee 
where $\alpha$ is a parameter that governs the low temperature behavior 
of the viscosity and $T_c(n)\simeq 0.167(3) T_F(n)$ is the critical 
temperature for the superfluid transition. This parametrization is
sufficiently flexible to accommodate the main possible behaviors 
of the shear viscosity at low temperature. For $\alpha>0$ the viscosity
tends to zero as $T\to 0$, for $\alpha \simeq 0$ the viscosity is 
approximately constant, and for $\alpha<0$ the viscosity diverges 
as $T\to 0$. We should note that the data mainly constrain the 
viscosity in the regime $0.5\lsim (T/T_c) \lsim 1.0$, and that the
value of $\alpha$ extracted from our analysis should not be taken
as a quantitative prediction for the shear viscosity at very low 
temperature, $(T/T_c)\lsim 0.5$.

We analyze the data in the superfluid regime using the results from
the previous section. As a first approximation we will solve the equation
using the equation of state in the superfluid regime, but assume that the 
normal and superfluid velocities are equal, $\vec{v}_n\simeq \vec{v}_s
\equiv \vec{v}$. We will then check this assumption by computing $\vec{w}=
\vec{v}_n-\vec{v}_s$ using equ.~(\ref{w_eom_lin}). We use the equation
of state and the initial state described in App.~\ref{sec_eos} and
\ref{sec_eos_trap}. The equations of fluid dynamics are solved using the 
anisotropic fluid dynamics method described in \cite{Bluhm:2017rnf,Bluhm:2015raa}.

\begin{figure}[t]
\centering
\includegraphics[scale=0.8]{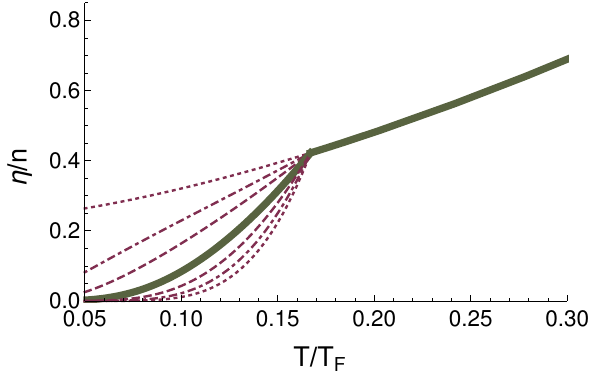}
\caption{\label{fig:eta-n}
Viscosity to density ratio $\eta/n$ as a function of $T/T_F$, where
$T_F$ is the local Fermi temperature. 
The lines show the fit obtained in this work, for different values of the 
parameter $\alpha$. The lines are labeled as in Fig.~\ref{fig:Ar-LowT}. In 
particular, the thick line corresponds to $\alpha=2$.}
\end{figure}

The results are shown in Fig.~\ref{fig:Ar-LowT}. We plot the data for 
$A_R(t^*)$ as a function of the initial central temperature $T/T_F$ of the 
cloud. Note that he critical temperature is $T_c/T_F=0.167(3)$. We also 
remark that in a trap, the superfluid initially appears in the center of 
cloud, and that near $T_c$ the size of the superfluid core is small (see
App.~\ref{sec_eos_trap} for an illustration of the superfluid and normal 
density profiles). As a consequence, we observe that in the regime $T/T_c
\in [0.8,1.0]$ the aspect ratio is only very weakly dependent on the viscosity
in the superfluid. The data show a noticeable change in slope of $A_R$ as a 
function of $T/T_F$ at lower temperatures, $T\lsim 0.8 T_c$. We find that 
a good description of the data in this regime can only be achieved if the 
viscosity at low temperature drops below the extrapolation from the normal 
phase. It is difficult to fully quantify this statement, because the data 
contain some outliers, and the hydrodynamic prediction for $A_R(t^*)$ is 
only weakly sensitive to the value of $\alpha$ beyond $\alpha\simeq 5$. 
Based on  Fig.~\ref{fig:Ar-LowT} we conclude that the data prefer $\alpha
\gsim 2$ (the prediction for $\alpha=2$ is shown as the thick green line 
in the figure). 

  The corresponding behavior of $\eta/n$ as a function of $T/T_F$ is shown
in Fig.~\ref{fig:eta-n}. We observe that the viscosity exhibits a fairly 
steep drop below the critical temperature. This behavior is in agreement 
with the reconstruction performed as part of the original experimental work 
\cite{Joseph:2014}. These authors find\footnote{
To be more precise, if we define $\alpha_n\equiv \eta/n$ as a function
of the ratio $(T/T_F)$ then $\alpha_n(0.8(T/T_F)_c) = 0.32 \alpha_n 
((T/T_F)_c)$.}
$\eta(0.8T_c)=(0.32\pm 0.22)\eta(T_c)$, compared to $\eta(0.8T_c)\simeq 
0.53\, \eta(T_c)$ from Fig.~\ref{fig:eta-n}. The analysis in 
\cite{Joseph:2014} is based on a number of simplifying assumptions. It
assumes, in particular, that there is a critical radial distance in the
expanding beyond which the cloud becomes free streaming. This radius is
adjusted to reproduce the theoretically known value of the shear viscosity
at large temperature \cite{Bruun:2005}. In our work the transitions to 
free streaming happens dynamically, governed by an extended hydrodynamic 
description that has been tested by comparison with exact numerical 
simulations of the Boltzmann equation \cite{Bluhm:2017rnf,Pantel:2014jfa}. 

\begin{figure}[t]
\centering
\includegraphics[width=0.45\hsize]{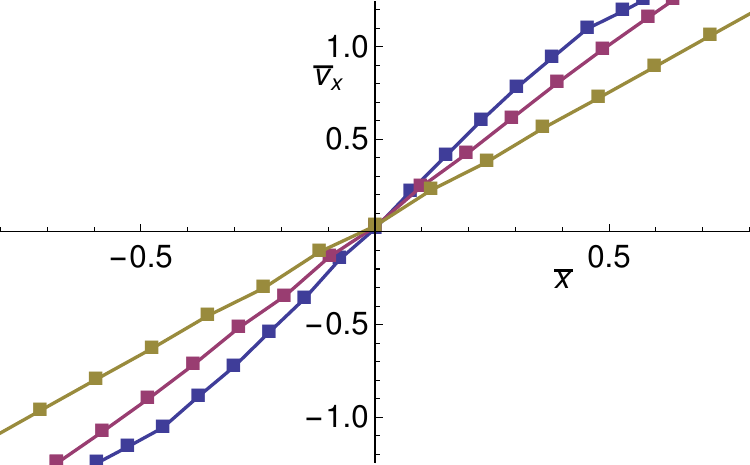}
\includegraphics[width=0.45\hsize]{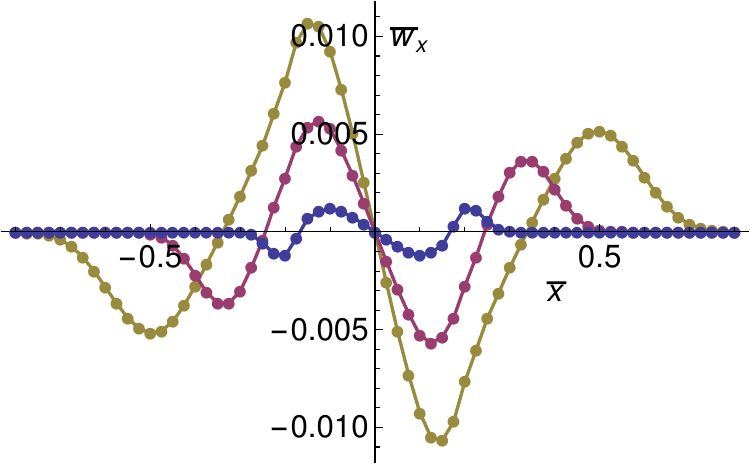}
\caption{\label{fig:w-pert}
Perturbative estimate of the superfluid velocity $\vec{w}=\vec{v}_n-\vec{v}_s$. 
The left panel shows the average velocity in the $x$-direction $v_x\simeq (v_n)_x 
\simeq (v_s)_x$ from a simulation of trapped gas with central temperature 
$(T/T_F)=0.094$. The position $\bar{x}=x/x_0$ and velocity $\bar{v}=v/v_0$ are 
given in dimensionless units, see  App.~\ref{sec_eos_trap}. The curves correspond 
to three different times $t=(0.25,0.50,0.75)\bar\omega^{-1}$ (blue, red, brown). 
The right panel shows the perturbative solution for $w_x$, as explained in the 
text. Note that the scale in the right panel is two orders of magnitude smaller.}
\end{figure}

 Finally, we can go beyond earlier work and test the consistency of the assumption 
that the normal and superfluid velocities are equal. For this purpose we solve
equ.~(\ref{w_eom_lin}) on the fluid dynamic background of the expanding solution 
with $\vec{w}=0$. The viscous heating rate is given by 
equ.~(\ref{e_cons},\ref{Q_diss})
\be
\label{visc_heat}
\dot{\cal E} = \frac{\eta}{2} \left(\sigma_{ij}\right)^2 \, , 
\ee
where $\sigma_{ij}$ is the strain tensor defined in equ.~(\ref{strain}).
If the the viscosity is small then the velocity is approximately linear 
in distance, and $\sigma_{ij}$ is spatially constant. This implies that 
spatial gradients in the rate of energy dissipation are mostly governed 
by the functional form of the shear viscosity. Note that a possible 
non-zero $\zeta_3$ does not contribute to energy dissipation in the limit
$\vec{w}\to 0$.

 The change in temperature is given by $\dot{T}=\dot{\cal E}/c_V$, where 
$c_V$ is the specific heat, see App.~\ref{sec_eos}. Note that $c_V$ drops
rapidly as $T/T_F\to 0$, so that the spatial structure of the temperature
profile is very sensitive to rate at which $\eta$ and $c_V$ approach their 
low temperature values. The equation of motion for $\vec{w}$ also involves
the ratio $s/\rho_n=s/(mn_n)$. Both the numerator and the denominator 
vanish in the limit $T\to 0$, but for the unitary Fermi gas the ratio
is $s/n_n$ close to unity and only weakly dependent on temperature, as 
shown in App.~\ref{sec_eos}. A numerical solution of equ.~(\ref{w_eom_lin}) 
is shown in Fig.~\ref{fig:w-pert}. We observe that spatial variations in 
dissipative heating do indeed generate a second sound-like perturbation. 
We note, however, that the amplitude of this perturbation is smaller than
the mean fluid velocity by almost two orders of magnitude. This means 
that the approximation $\vec{w}=0$ is consistent.

\section{Summary and Outlook}
\label{sec_outlook}

 In this work we have summarized the formalism for applying superfluid
hydrodynamics to the dilute Fermi gas at unitarity, and constructed a
suitable equation of state. We have showed that, unless the viscosity is 
very large, the equations can be solved perturbatively in the variable
$\vec{w}=\vec{v}_n-\vec{v}_s$. We have numerically studied the evolution
of a trapped Fermi gas after release from a deformed trap in the regime
where the core of the cloud is superfluid. Comparing our simulations
to the experimental data obtained in \cite{Joseph:2014} we conclude that 
the viscosity must drop significantly below $T_c$. This drop can be 
parametrized as an exponential decrease proportional to $\exp(-\alpha 
(T_c/T))$, where $\alpha\gsim 2$.

 Our results are in some tension with the data obtained in 
\cite{Patel:2019}. Patel et al.~measure the sound attenuation in a
unitary Fermi gas confined in a box trap at approximately constant
density and different temperatures, both above and below $T_c$. They
find that the sound diffusivity is approximately constant around
$T_c$, and does not exhibit a pronounced decrease. The sound attenuation
constant involves several transport coefficients, including the shear
viscosity, the thermal conductivity, and, below $T_c$, the bulk 
viscosity coefficient $\zeta_3$. These transport coefficient can
be disentangled using linear response measurements, as demonstrated
in \cite{Baird:2019}, but this analysis has not been performed below
$T_c$. This implies that it is possible that the sound diffusivity
remains constant despite the fact that the viscosity is decreasing, 
but this behavior appears unlikely, and is not predicted by any 
transport theory analysis available in the literature. 

 It seems more plausible that the difference is explained by 
differences in the experimental approach. For example, it is possible
that the viscosity of the superfluid phase is governed by a fairly
dilute gas of quasi-particles, whereas transport in the normal phase
is controlled by a system of dense, strongly correlated, excitations.
In this case the superfluid core of an expanding gas cloud might be
too small to exhibit dissipative two-fluid dynamics. This option 
can be addressed by studying linear response in a box trap, and 
we plan to extend our analysis to these systems in the future. 
The breakdown of superfluid dissipative hydrodynamics is observed
in the collective mode experiments \cite{Patel:2019}. Dissipative 
fluid dynamics predicts that sound wave damping scales as wave 
number squared. This is seen over a wide range of wave numbers
above $T_c$, but only in a much narrower window below $T_c$.

The work of J.~H. and T.~S.~was supported in part by the US Department of 
Energy grant DE-FG02-03ER41260. We thank J.~Thomas and M.~Zwierlein for
many useful discussions, and for providing us with the data published
in \cite{Ku:2011} and \cite{Joseph:2014}. 

\appendix
\section{Galilean transformations}
\label{sec_gal}

Under a Galilean transformation with boost velocity $\vec{u}$
we have
\bea
\label{rho_boost}
\rho'   &=& \rho \, , \\
\label{E_boost}
{\cal E}'   & =& {\cal E} +\vec\jmath\cdot\vec{u}
+ \frac{1}{2}\rho u^2  \, , \\
\label{j_boost}
\jmath'_i &=& \jmath_i +\rho u_i\, ,  \\
\label{Q_boost}
 Q_i' &=& Q_i + u_j \Pi_{ij} + \frac{1}{2}u^2 \jmath_i  
   +{\cal E}' u_i\, , \\
\label{Pi_boost}
\Pi_{ij}' &=& \Pi_{ij} + u_i\jmath_j+u_j\jmath_i + u_iu_j \rho\, , 
\eea
where $\rho$ is the mass density, ${\cal E}$ is the energy density,
$\vec{\jmath}$ is the mass current (momentum density), $\vec{Q}$ is 
the energy current, and $\Pi_{ij}$ is the stress tensor.
The pressure and the entropy density of the fluid are Galilean invariant
\be
 P'=P, \hspace{1cm} s'=s\, .
\ee
In a superfluid the normal and superfluid densities are separately 
Galilean invariant. The fluid velocities transforms as
\be
\vec{v}_\alpha^{\, \prime} = \vec{v}_\alpha+\vec{u}\, , 
\ee
where $\alpha=(n,s)$.

\section{Superfluid hydrodynamics in the normal fluid rest frame}
\label{sec_sfl_nfl}

 In Section \ref{sec_sfl_Landau} we studied the thermodynamics of 
a moving superfluid by constructing the energy density in the superfluid  
rest frame. This procedure is well defined, even in the limit $T\to T_c$. 
However, one may be concerned that this method is not the best choice 
in the limit that the superfluid density is much smaller than the total 
density of the fluid. In this appendix we consider an alternative approach 
based on thermodynamic identities in the rest frame of the normal fluid. 
Consider the total energy density of a superfluid. Following \cite{Ho:1998}
we write
\be 
\label{dE}
d{\cal E}  = \mu_{\jmath} dn + Tds + \vec{v}_n\cdot d\vec\jmath 
  + \vec\jmath_n\cdot d\vec{v}_s \, . 
\ee
which defines $\vec{v}_n$ and $\vec{\jmath}_n$ as variables conjugate to the 
momentum density $\vec{\jmath}$ and the superfluid velocity $\vec{v}_s$. We 
use the notation $\mu_{\jmath}$ to indicate that the chemical potential is 
defined at fixed $\vec\jmath$. Using Galilean invariance,  ref.~\cite{Ho:1998} 
shows that $\vec\jmath_n=\vec\jmath-\rho \vec{v}_n$, so that $\vec\jmath_n$ 
is the current in the normal fluid rest frame. We can also write $\vec\jmath
=\rho_n \vec{v}_n + \rho_s \vec{v}_s$, so that $\vec\jmath_n=\rho_s(\vec{v}_s
-\vec{v}_n)\equiv -\rho_s \vec{w}$. We obtain the pressure by performing a 
Legendre transformation, 
\be 
P=-{\cal E} +\mu_\jmath n + Ts+\vec{v}_n\cdot\vec\jmath
\ee
so that 
\be 
dP  = n d\mu_{\jmath}  + sdT +\vec\jmath\cdot d\vec{v}_n
 -\vec{\jmath}_n\cdot d \vec{v}_s \, . 
\ee
Using the explicit form of the currents in terms of the normal 
and superfluid densities as well as velocities we find the Gibbs-Duhem 
relation
\be 
\label{P_GD_nfl}
 dP = n\,d\mu_n + s\, dT - \frac{\rho_s}{2} dw^2\, ,
\ee
where we have defined the chemical potential in the normal fluid frame
\be 
\mu_n = \mu_\jmath + \frac{1}{2} mv_n^2\, . 
\ee
This Gibbs-Duhem relation implies that the superfluid density can be 
defined as 
\be 
\label{rho_s_def}
 \rho_s = -2 \left.\frac{\partial P}{\partial w^2}\right|_{\mu_n,T}
\ee
This result should be compared with equ.~(\ref{rho_n_def}). We note that
the dependence of $P$ on $\vec{w}$ depends crucially on what chemical
potential, $\mu_n$ or $\mu_s$, is held constant. The energy density 
in the normal fluid frame can be obtained via a Galilei transformation. 
We have 
\be 
 {\cal E}_n = {\cal E} -\vec\jmath\cdot \vec{v}_n + \frac{1}{2}\rho v_n^2
\ee
which implies that the pressure can be written as 
\be 
\label{P_nfl}
 P = -{\cal E}_n + \mu_n n +Ts \, . 
\ee
Note that in the normal fluid frame we have the usual (one-fluid) 
relation $P+{\cal E}_n=\mu_n n+sT$. 

In a scale invariant Fermi gas we can write
\be
\label{P-uni-1}
P(\mu_n,T,\vec{w}) = m^{3/2}\mu_n^{5/2}
F_n\left( \frac{T}{\mu_n},\frac{mw^2}{\mu_n}\right)\, . 
\ee
As before, we can try to simplify the problem by expanding the 
pressure in $\vec{w}$, 
\be
\label{P-uni-2}
P(\mu_n,T,\vec{w}) = 
      m^{3/2}\mu_n^{5/2} p_{n0}\left( \frac{T}{\mu_n}\right)
    - m^{5/2}\mu_n^{3/2} w^2 p_{n1}\left(\frac{T}{\mu_n}\right)\, . 
\ee
In the limit $\vec{w}\to 0$ this function must agree with 
equ.~(\ref{P-uni-4}) so that $p_{n0}(s)=p_{s0}(x)=p(x)$. The second 
function $p_{n1}(x)$ determines the superfluid mass density
\be
\rho_s = 2m^{5/2} \mu_0^{3/2}  p_{n1}\left(\frac{T}{\mu_n}\right).
\ee
which has been measured in \cite{Grimm:2013,Patel:2019}. Note that 
$\rho_s$ is positive, so the term proportional to $w^2$ lowers the 
pressure at fixed $\mu_n$ and $T$. The energy density in the normal 
fluid rest frame can be determined using
\be
 {\cal E}_n = \left\{ \mu_n\frac{\partial}{\partial \mu_n}
 + T\frac{\partial}{\partial T} - 1 \right\} P
\ee
where $P(\mu_n,T,w)$ is given in equ.~(\ref{P-uni-2}) above. At $O(w^2)$
we find that
\be
\label{P-E-n}
P = \frac{2}{3}{\cal E}_n -\frac{1}{3}\rho_s w^2 \, . 
\ee
At this order we also find that 
\be
\label{P-E-n-2}
P = \frac{2}{3} \left\{ {\cal E} - \frac{1}{2}\rho_n v_n^2
- \frac{1}{2}\rho_s v_s^2 
  \right\}\, . 
\ee
Like the results in Sect.~\ref{sec_sfl_Landau} the expressions in
equ.~(\ref{P-E-n}) and (\ref{P-E-n-2}) are not given in terms of 
the primary variables $({\cal E},\rho,\vec\jmath, \vec{v}_s)$, 
and the equation of state is needed to determine $\rho_n,\rho_s$ 
and $\vec{v}_n$. This can be accomplished by tabulating the equation 
of state, or by solving for $\rho_s$ iteratively in $\vec{w}$.

\begin{figure}[t]
\centering
\includegraphics[scale=0.8]{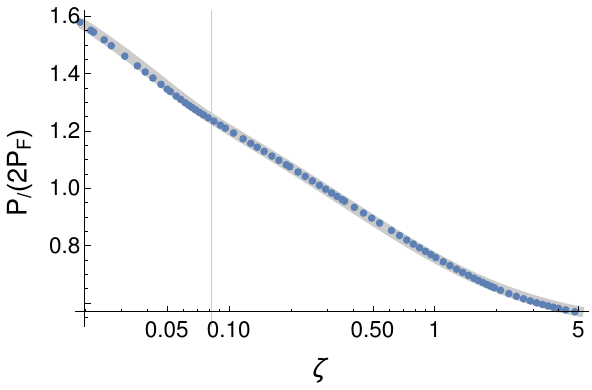}
\caption{\label{fig:h}
Pressure $P(\zeta)$ as a function of inverse fugacity 
$\zeta=\exp(-\mu/T)$ in units of the pressure $P_F$ of a free 
Fermi gas. The blue dots show the data taken by the MIT group. 
The solid line is our parametrization, which has a discontinuity
at $\zeta=\zeta_c=0.082$.}
\end{figure}

\section{Equation of state}
\label{sec_eos}

 In this appendix we describe a parametrization of the equation 
of state of the unitary Fermi gas. We follow the basic strategy 
described in App.~B of \cite{Bluhm:2017rnf}, but we extend the 
method to the regime below the critical temperature. We begin by
considering the pressure of a Fermi gas at rest. We write the 
pressure as\footnote{Note that the function $p(x)$ defined in 
equ.~(\ref{P-uni-0}) is given by $p(x)=(2\pi)^{-3/2}x^{5/2}
f(\exp(-1/x))$.}
\begin{equation}
\label{f_def}
P(T,\mu)=T\lambda^{-3} f(\zeta)\, , 
\end{equation}
where $\lambda=[2\pi/(mT)]^{1/2}$ is the thermal de Broglie wave length,
and $\zeta=exp(-\mu/T)$ is the inverse fugacity. In the regime above 
the critical temperature $T_c$ it is useful to represent the function
$f(\zeta)$ in terms of the result for a free Fermi gas
\begin{eqnarray}\label{pressure}
f(\zeta)=h(\zeta)p_F(\zeta),\quad 
p_F(T,\mu)=-Li_{5/2}(-\zeta^{-1}).
\end{eqnarray}
The function $h(\zeta)$ was measured in \cite{Ku:2011}, see Fig.~\ref{fig:h} 
We follow our previous work and parametrize $h(\zeta)$ in the normal fluid 
regime by a Pade approximant
\begin{equation}
\label{h_z_pade}
\frac{h(\zeta)}{2}=\frac{\zeta^2+c_1 \zeta+c_2}{\zeta^2+c_3 \zeta +c_4},
\hspace{1cm} \left( \zeta > \zeta_c = 0.082\right)
\end{equation}
with
\begin{equation}
c_1=1.32109, \quad c_2=0.026341, \quad c_3=0.541993, \quad c_4=0.005660.
\end{equation}
Here, $\zeta_c=0.082$ is the critical value of the inverse fugacity
obtained in \cite{Ku:2011}. Once the pressure is given, other 
thermodynamic observables are easily determined. We can write the 
density and entropy density as
\begin{figure}[t]
\begin{center}
\includegraphics[scale=0.8]{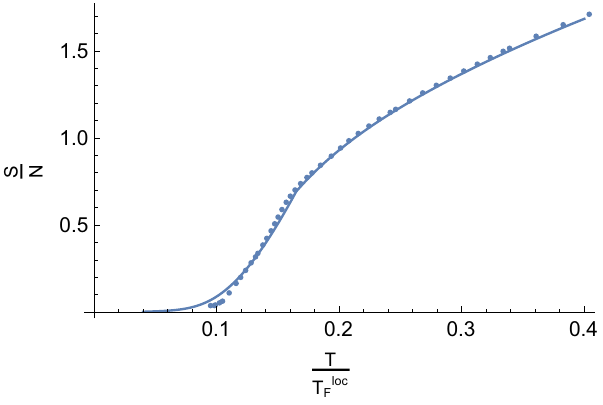}
\end{center}
\caption{\label{fig:S_N}
Entropy per particle $S/N$ as a function of temperature
in units of Fermi temperature $T/T_F$. Here, $T_F$ is the local 
Fermi temperature of the gas, defined by $k_B T_F=k_F^2/(2m)$ and
$k_F^3=3\pi^2 n$. The data points are from \cite{Ku:2011}.}
\end{figure}
\be
\label{g_def}
n(\mu,T)= \lambda^{-3} g(\zeta)\, , \hspace{0.5cm}
s(\mu,T) = \lambda^{-3} k(\zeta)\, . 
\ee
where 
\begin{eqnarray}
g(\zeta)&=&-Li_{3/2}(-\zeta^{-1})h(\zeta)+\zeta Li_{5/2}(-\zeta^{-1})
h^{\prime}(\zeta)\, ,  \\
k(\zeta) &=&  -\left(\log(\zeta){\it Li}_{3/2}(-\zeta^{-1}) 
                  +\frac{5}{2}{\it Li}_{5/2}(-\zeta^{-1}) \right)h(\zeta)
    +\zeta\log(\zeta){\it Li}_{5/2}(-\zeta^{-1})h'(\zeta)\, .
\end{eqnarray}
Other thermodynamic functions can be computed by taking additional 
derivatives. For example, the specific heat is given by 
\be
 c_{V} = \frac{T}{V}\left.\frac{\partial S}{\partial T}\right|_{V}
 = T\left[\left.\frac{\partial s}{\partial T}\right|_{\mu}
          -\frac{[(\partial n/\partial T)|_{\mu}]^2}
                {(\partial n/\partial \mu)|_{T}}\right]\, .
\ee
The parametrization in equ.~(\ref{h_z_pade}) is quite accurate, even
at temperatures below $T_c$. However, $T=T_c$ is a genuine critical
point and the parametrization should exhibit a non-analyticity at 
$\zeta=\zeta_c$. Furthermore, equ.~(\ref{h_z_pade}) does not 
very accurately describe derivatives of the pressure in the regime 
$\zeta<\zeta_c$. In particular, the entropy density is unphysical
for small values of $\zeta$.

 In order to address these issues we employ a separate fit of the 
pressure in the regime $\zeta<\zeta_c$. We have chosen a physically 
motivated model of the pressure which is of the form 
\begin{equation} 
\label{PlowT}
P(\mu,T)=\frac{2^{3/2}\mu^{5/2} m^{3/2}}{15\pi^2 \xi^{3/2}}
+ \frac{\pi^2 T^4 }{90}\left( \frac{3m}{2\mu}\right)^{3/2} 
 +A \mu^{5/2}m^{3/2}\sqrt{\frac{T}{\mu}}e^{-\frac{\mu B}{T}}.
\end{equation}
Here, the first term is the zero temperature pressure expressed in
terms of the Bertsch parameter $\xi=0.376$. The second term is the 
contribution of phonons in the superfluid phase, see, for example, 
ref.~\cite{Braby:2010ec}. The third term takes into account thermally
excited fermionic quasi-particles, where $A$ and $B$ are treated 
as fit parameters. The structure of this term is taken from mean field
calculations of the pressure in the BCS limit, see \cite{Gorkov}.
We fix the values of $A$ and $B$ by requiring the pressure and 
density to be continuous (but not differentiable) at $\zeta=\zeta_c$. 
This procedure ensures that the entropy density is continuous as
well. We obtain
\begin{equation}
A=4.6699,\quad \quad \quad B=2.15436.
\end{equation}
Note that we have not attempted to reproduce the exact critical 
behavior of the equation of state, which are expected to be those
of the three dimensional $O(2)$ model. In a harmonic trap, the 
critical region is only a narrow shell in coordinate space, and
critical behavior is difficult to observe. 

\begin{figure}[t]
\begin{center}
\includegraphics[scale=0.8]{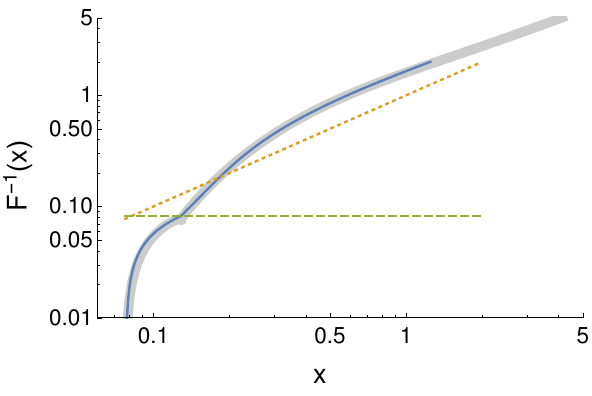}
\end{center}
\caption{\label{fig:Finv}
This plot shows the function $\zeta=F^{-1}(x)$ that determines
the inverse fugacity in terms of a dimensionless ratio of the 
pressure and the density, $x\sim (mP)^{3/2}/n^{5/2}$. The solid blue
line shows the two-component fit to the pressure and density. The 
dashed horizontal line is $\zeta_c=0.082$ and orange dotted line is
the high temperature limit $F^{-1}(x)\simeq x$. The grey band is 
a fit to $F^{-1}(x)$ described in the text.}
\end{figure}

  In terms of the functions $f(\zeta)$ and $g(\zeta)$ defined 
in equ.~(\ref{f_def}) and (\ref{g_def}) the low temperature
model of the equation of state is
\begin{eqnarray}
\label{f_s}
f_s (\zeta) &=&
\frac{2^{5/2} (2\pi)^{3/2}}{15\pi^2\xi^{3/2}} 
  \left( -\log(\zeta) \right)^{5/2}
+ \frac{\pi^2(2\pi)^{3/2}}{90} \left(\frac{3}{2}\right)^{3/2}
    \left( -\log(\zeta) \right)^{-3/2} \nonumber \\
 && \mbox{}+  A(2\pi)^{3/2} \left(\log(\zeta)\right)^2 \zeta^B \\
\label{g_s}
 g_s(\zeta) &=& \frac{2^{5/2} (2\pi)^{3/2}}{6\pi^2\xi^{3/2}} 
  \left( -\log(\zeta) \right)^{3/2}
- \frac{\pi^2(2\pi)^{3/2}}{90} \left(\frac{3}{2}\right)^{5/2}
    \left( -\log(\zeta) \right)^{-5/2} \nonumber \\
     &&  \mbox{}
     - A (2\pi)^{3/2} \left(2+B\log(\zeta)\right)
       \log(\zeta)  \zeta^B
\end{eqnarray}
where $f_s(\zeta)=f(\zeta<\zeta_c)$ and $g_s(\zeta)=g(\zeta<\zeta_c)$.
Equ.~(\ref{f_def}-\ref{h_z_pade}) and (\ref{f_s}) define our
equation of state. To illustrate the accuracy of this parametrization
we show in Fig.~\ref{fig:h} the pressure as a function of $\zeta$
and in Fig.~\ref{fig:S_N} the entropy per particle as a function 
of $T/T_F$, both compared to the experimental results of the MIT
group. 

\begin{figure}[t]
\begin{center}
\includegraphics[scale=0.8]{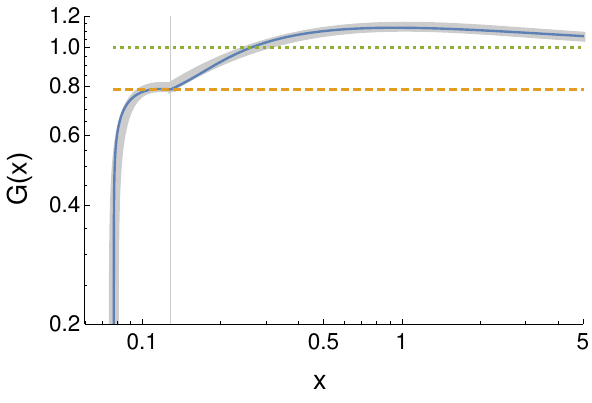}
\end{center}
\caption{\label{fig:Gcor}
Temperature correction factor $G(x)=P/(nT)$ as a function of the 
dimensionless variable $x\sim (mP)^{3/2}/n^{5/2}$. The solid blue
line shows the two-component fit to the pressure and density. The 
short dashed horizontal line is the high temperature limit $G=1$, 
and the long dashed line shows the critical value $G_c$. The grey band is 
a fit to $G(x)$ described in the text.}
\end{figure}

 In fluid dynamics we have to reconstruct $\zeta$ from the density 
and pressure of the gas. For this purpose we consider the function
\begin{equation}
F(\zeta)=\frac{2 f(\zeta)^{3/2}}{g(\zeta)^{5/2}}, 
\end{equation}
where $f(\zeta)$ and $g(\zeta)$ are defined piecewise for $\zeta$
larger and smaller than $\zeta_c$. The function $F(\zeta)$ is 
proportional to the dimensionless ratio $(mP)^{3/2}/n^{5/2}$.
We have 
\begin{equation}
\zeta=F^{-1}\left( 
\frac{2}{(2 \pi)^{3/2}} \frac{(mP)^{3/2}}{n^{5/2}}\right).
\end{equation}
The function $F(\zeta)$ is defined so that $F(\zeta\gg\zeta_c)\simeq
\zeta$, and as a result $F^{-1}(x\gg x_c)\simeq x$. We show $F^{-1}(x)$
for all $x$ in Fig.~\ref{fig:Finv}. Note that there is a minimum value
of $x$, given by $x_0=F(0)$. In practice, we employ parametrization of 
$F^{-1}(x)$. This function is also defined piecewise for $x>x_c$ 
(corresponding to $T>T_c$) and $x<x_c$ (the regime $T<T_c$), 
where $x_c=0.1285$. For $x>x_c$ we again use a Pade approximant
\be
\label{Finv_pade}
F^{-1}_{\it fit}(x)=x \, \frac{1+h_1 /x+h_2/x^2}{1+h_3/x +h_4/x^2} ,
\hspace{1cm} \left( x > x_c = 0.1285\right)
\ee
with
\be
h_1= 1.1601  \, ,  \quad
h_2= -0.0927 \, , \quad
h_3= 0.2119 \, , \quad
h_4= 0.07729\, .  
\ee
In the superfluid regime $x<x_c$ we write
\be 
F^{-1}_{\it fit}(x) = \zeta_c^{\it fit} 
   \left\{ 1 -\left( \frac{x_c-x}{x_c-x_0}\right)^{3/2}\right\}^{2/3}
\ee
with
\be 
 \zeta_c^{\it fit} = 0.07732 \, , \quad
 x_c = 0.1285 \, , \quad 
 x_0 = 0.0775 \, . 
\ee
The function $F^{-1}(x)$ together with the fit given above is shown
in Fig.~\ref{fig:Finv}.

\begin{figure}[t]
\begin{center}
\includegraphics[scale=0.8]{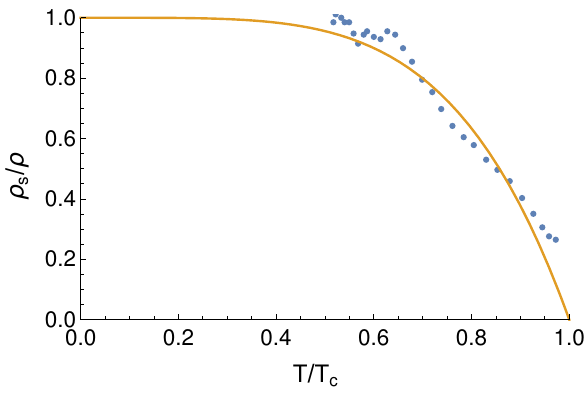}
\end{center}
\caption{\label{fig:rhos}
Superfluid mass fraction $\rho_s/\rho$ as a function of the 
temperature $T$ in units of the local Fermi temperature $T_F$.
The points show the data from \cite{Grimm:2013}, and the line
shows the fit discussed in the text. Note that the data points
have uncertainties of order 10\%, both in $T/T_F$ and $\rho_s/\rho$,
that are not shown here.}
\end{figure}

  Finally, given the local fugacity we have to determine the 
temperature and chemical potential of the fluid. In the high 
temperature limit this is straightforward; we can use $T=P/n$.
In the general case we can write
\begin{equation}
T=G(x)\frac{P}{n} \, , 
\end{equation}
where $G(x)$ is a correction factor, given by 
\be
G(x)=\frac{g(\zeta(x))}{f(\zeta(x))}
  =\frac{g(F^{-1}(x))}{f(F^{-1}(x))}\, .
\ee
The function $G(x)$ extracted from our parametrization of the 
pressure and density is shown in Fig.~\ref{fig:Gcor}. We note 
that for $x>x_c$ the function $G(x)$ is close to the high 
temperature limit $G=1$. In the low temperature regime $x<x_c$
the correction factor $G(x)$ drops very steeply, with $G(x_0)=0$
at $x_0=0.0775$. This behavior motivates a simple two-component
fit, similar to the one for $F^{-1}(x)$. We write
\be
\label{Gcor_pade}
G_{\it fit}(x)= \frac{1+d_1 /x+d_2/x^2}{1+d_3/x +d_4/x^2} ,
\hspace{1cm} \left( x > x_c = 0.1285\right)
\ee
with
\be
d_1= 1.8052  \, ,  \quad
d_2=-0.0022  \, , \quad
d_3= 1.3668  \, , \quad
d_4= 0.1179  \, .  
\ee
In the superfluid regime $x<x_c$ we write
\be 
G_{\it fit}(x) = G_c^{\it fit} 
   \left\{ 1 -\left( \frac{x_c-x}{x_c-x_0}\right)^{4}\right\}^{1/4}\, . 
\ee
with $G_c^{\it fit} =0.7944.$

\begin{figure}[t]
\begin{center}
\includegraphics[scale=0.8]{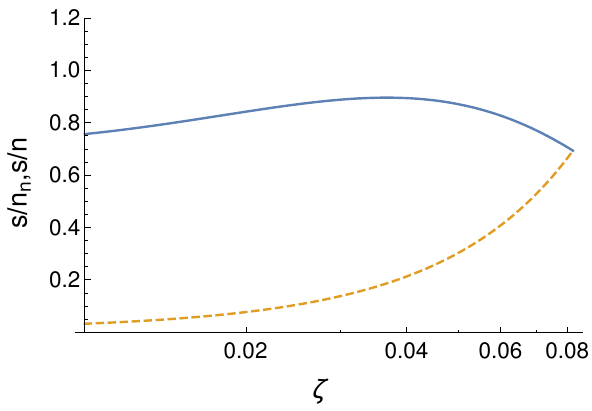}
\end{center}
\caption{\label{fig:SoNn}
Entropy density over superfluid density (solid line), and 
entropy density over total density (dashed line) in the superfluid 
regime, plotted as a function of the inverse fugacity $\zeta$.}
\end{figure}

 As explained in the text, in the superfluid phase there is an 
additional thermodynamic function, the superfluid mass density
$\rho_s(\mu,T)$. The superfluid mass fraction was determined in
\cite{Grimm:2013}, see also the recent work \cite{Patel:2019}. We 
show the results of \cite{Grimm:2013} in Fig.~\ref{fig:rhos}. A 
simple quasi-particle model for these results is discussed by Baym 
and Pethick \cite{Baym:2013}. Here we use an even simpler 
parametrization, given by
\be
\frac{\rho_s}{\rho} = 1-\left(\frac{T}{T_c}\right)^{9/2}\, , 
\ee
where $T_c$ is the critical temperature. This parametrization is 
not directly motivated by a physical model, but numerically close 
to the theory of Baym and Pethick. Note that the superfluid density 
of a dilute Bose gas scales as $\rho_s/\rho=1-(T/T_c)^{3/2}$.
Finally, based on this result we can compute the ratio of the entropy
density over the normal fluid density, which enters in the equation 
for the acceleration of $\vec{w}$, see equ.~(\ref{w_eom_lin}). The
result is shown in Fig.~\ref{fig:SoNn}.

\begin{figure}[t]
\begin{center}
\includegraphics[scale=0.8]{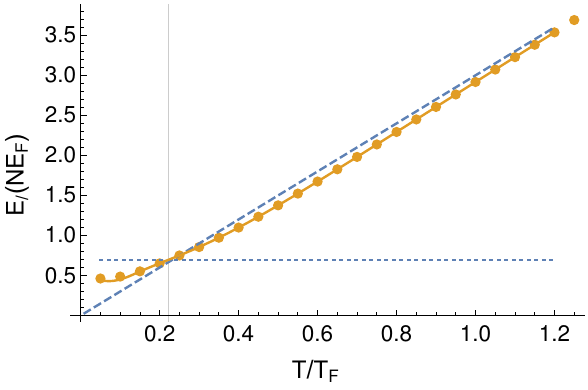}
\end{center}
\caption{\label{fig:EoEF}
Energy versus temperature for a harmonically trapped Fermi gas.
The energy is given in units of $NE^{\it trap}_F$, where $E^{\it trap}_F 
=3N^{1/3}\bar\omega$. The temperature is given in units of $T^{\it trap}_F 
=E^{\it trap}_F$ (where $k_B=1$). The points are computed from our 
parametrization of the equation of state, the dotted line is an
interpolating function. The dashed diagonal line corresponds to 
the high temperature limit $E=(3/2)T$. The horizontal and vertical 
lines indicate the critical values $T_c/T^{\it trap}_F=0.222$ and
$E_c/E^{\it trap}_F=0.695$.}
\end{figure}

\section{Trapped Fermi gas}
\label{sec_eos_trap}
 
\begin{figure}[t]
\begin{center}
\includegraphics[scale=0.8]{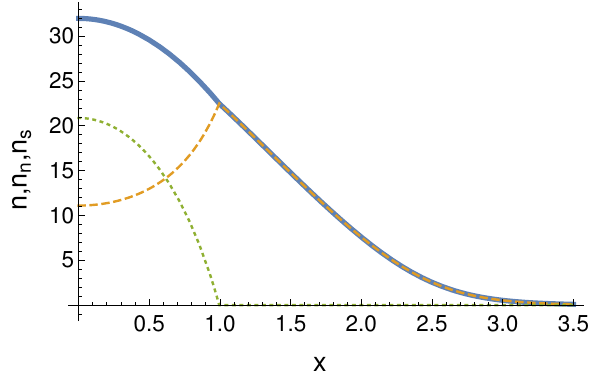}
\end{center}
\caption{\label{fig:Profile}
Density profile of a harmonically trapped Fermi gas below the 
critical temperature at which superfluidity appears at the center
of the trap. We show the total (blue), normal (yellow, dashed), and
superfluid (green, dotted) densities $n,n_n$ and $n_s$ in units of
$\lambda^{-3}$ as a function of $x$ in units of the length $x_0$ defined
in the text. Here, the central inverse fugacity was chosen as
$\zeta_0=0.05$.}
\end{figure}

   A trapped Fermi gas in thermal equilibrium is a solution of 
the hydrostatic equation. As explained in Sect.~\ref{sec_sol}
in both the normal and the superfluid regime the solution is given by
\be 
\label{n_LDA_app}
 n(\vec{x},t) = n(\mu_s(\vec{x}),T)\, \hspace{1cm}
 \mu_s(\vec{x})=\mu_c - V_{\it ext}(\vec{x})\, . 
\ee
In the experiment we consider trapped clouds with a given 
number of particles $N$ at a fixed temperature $T$ or total
energy $E$. Given $N$ and $T$ the central inverse fugacity 
$\zeta_0=\zeta(0)$ is fixed by the condition
\be
\label{zetacenter}
\frac{3}{(2 \pi)^3} \left( \frac{T}{T^{\it trap}_F}\right) 
\int d^3 x \, g \left( \zeta_0 \exp \left( \frac{x^2}{2}\right)\right) 
\equiv 1\, , 
\ee
where $T^{\it trap}_F=3N^{1/3}\bar{\omega}$ is the Fermi temperature 
of the trap. Once $\zeta_0$ is determined the total energy can be 
computed using the virial theorem. Making use of the virial theorem 
we can calculate the energy of the trapped gas
\be
\label{energyini}
\frac{E}{NE^{\it trap}_F}=\left(\frac{T}{T^{\it trap}_F}\right)^3 
  \frac{\int d^3 x\, x^2 g \left( \zeta_0 \exp \left(\frac{x^2}{2} 
  \right)\right)}
  {\int d^3x \, g \left( \zeta_0 \exp \left(\frac{x^2}{2} \right)\right)} \, .
\ee
Fig.~\ref{fig:EoEF} shows $E/(NE^{\it trap}_F)$ as a function of $T/T^{\it trap}_F$ 
for a harmonically trapped Fermi gas. The critical temperature and energy, 
that means the value of $T$ and $E$ at which superfluidity appears at 
the center of the trap, are 
\be 
T_c/T^{\it trap}_F = 0.222\, , \quad
E_c/(NE^{\it trap}_F) = 0.695
\ee
The zero temperature limit of the energy is $E_0/(NE^{\it trap}_F)=
(3\sqrt{\xi})/4 =0.460$.

 An example for the density profile of a harmonically trapped Fermi gas 
in the superfluid regime is shown in Fig.~\ref{fig:Profile}. In this example 
the inverse fugacity at the trap center is $\zeta_0=0.05$, corresponding to 
a central temperature $T/T_F=0.13$. We note that there is a two-fluid 
mixture in the core. The superfluid appears at some critical radius
$x_c$, but the total density only shows a very mild non-analyticity 
at $x_c$. In the figure the position $x$ is shown in dimensionless 
units $\bar{x}=x/x_0$ with 
\be 
 x_0 = 
    \left[ \frac{2}{3} \frac{(3N)^{1/3}}{m\bar{\omega}}\right]^{1/2}\, .  
\ee
Similar dimensionless can be employed for time $\bar{t}=t/t_0$ with 
$t_0=\bar{\omega}^{-1}$ and velocity $\bar{v}= v/v_0$ with $v_0=t_0/x_0$. 
In our hydrodynamic simulations we also use dimensionless variables for 
thermodynamic quantities, such as density $\bar{n}=n/n_0$, pressure $\bar{P}
=P/P_0$, temperature $\bar{T}=T/T_0$, and viscosity $ \bar{\eta}=\eta/\eta_0$
\be 
 n_0=x_0^{-3}, \quad
 P_0=m\bar{\omega}^2x_0^{-1}, \quad
 T_0=m\bar{\omega}^2 x_0^2, \quad
 \eta_0 = m\bar{\omega}x_0^{-1}. 
\ee


\end{document}